# Effect of Coulomb impurities on the electronic structure of magic angle twisted bilayer graphene

Muhammad Sufyan Ramzan,[1,2,3] Zachary A. H. Goodwin, [4,5,6] Arash A. Mostofi, [4] Agnieszka Kuc, [1,3] and Johannes Lischner[4*]

[1]*Department of Physics and Earth Sciences, Jacobs University Bremen, Campus Ring 1, 28759 Bremen, Germany*

[2]*Institut für Physik, Carl von Ossietzky Universität Oldenburg, 26129 Oldenburg*

[3]*Helmholtz-Zentrum Dresden-Rossendorf, Abteilung Ressourcenökologie, Forschungsstelle Leipzig, Permoserstr. 15, 04318 Leipzig, Germany*

[4]*Departments of Materials and Physics and the Thomas Young Centre for Theory and Simulation of Materials, Imperial College London, South Kensington Campus, London SW7 2AZ, UK*

[5]*National Graphene Institute, University of Manchester, Booth St. E. Manchester M13 9PL, United Kingdom*

[6]*School of Physics and Astronomy, University of Manchester, Oxford Road, Manchester M13 9PL, United Kingdom*

*E-Mail: j.lischner@imperial.ac.uk

In graphene, charged defects break the electron-hole symmetry and can even give rise to exotic collapse states when the defect charge exceeds a critical value which is proportional to the Fermi velocity. In this work, we investigate the electronic properties of twisted bilayer graphene (tBLG) with charged defects using tight-binding calculations. Like monolayer graphene, tBLG exhibits linear bands near the Fermi level but with a dramatically reduced Fermi velocity near the magic angle (approximately 1.1°). This suggests that the critical value of the defect charge in magic-angle tBLG should also be very small. We find that charged defects give rise to significant changes in the low-energy electronic structure of tBLG. Depending on the defect position in the moiré unit cell, it is possible to open a band gap or to induce an additional flattening of the low-energy valence and conduction bands. Our calculations suggest that the collapse states of the two monolayers hybridize in the twisted bilayer. However, their in-plane localization remains largely unaffected by the presence of the additional twisted layer because of the different length scales of the moiré lattice and the monolayer collapse state wavefunctions. These predictions can be tested in scanning tunnelling spectroscopy experiments.

## 1. Introduction

Since the experimental discovery of correlated insulator states and unconventional superconductivity in magic-angle twisted bilayer graphene (tBLG),[1,2] there has been significant interest in understanding the electronic properties of this system.[3–22] The low-energy electronic structure of tBLG is characterized by a set of four bands, which become extremely flat as the twist angle approaches the magic angle of approximately 1.1°. [23–29] Near the K and K' points of the moiré Brillouin zone, these bands exhibit a linear dispersion – similar to monolayer graphene – but with a dramatically reduced Fermi velocity. [23–25,30,31]

To date, most experimental and theoretical works have focused on understanding the properties of pristine tBLG, but it is well known that defects play an important role in real devices that exploit the properties of two-dimensional (2D) materials.[11,32] In particular, impurities that donate electrons or holes can be used to control the concentration of charge carriers.[33–39] Once ionized, these impurities act as Coulomb scatterers and reduce the charge carrier mobility.[40] Recently, charged defects in tBLG have been investigated by Larson et al.[41] using *ab initio* density-functional theory (DFT). They showed that intercalated lithium atoms reside in regions of AA-stacking in the moiré unit cell and act as electron donors. Interestingly, they also found that the intercalation gives rise to a slight deformation of the band structure. However, only relatively large twist angles (7.34° and 2.45°) were studied[42,43] and this motivates our study of charged defects in tBLG near the magic angle.

Charged defects have been extensively studied in monolayer graphene.[44–47] As electrons in graphene obey a Dirac equation like relativistic particles (but with the Fermi velocity of graphene replacing the speed of light), their response to a charged defect depends sensitively on the magnitude of the defect charge. Specifically, it has been found that there is a critical value of the defect charge, given by $Z_c = \epsilon v_F/2$ (in atomic units), where $\epsilon$ is the effective dielectric constant (containing both the internal screening in the graphene and the external screening due to the substrate) and $v_F$ is the Fermi velocity. [48,49] When the defect charge is smaller than $Z_c$, the system is in the so-called subcritical regime characterized by a broken electron-hole symmetry. When the system is in the supercritical regime, i.e., when the defect charge exceeds $Z_c$, a novel collapse state is formed which gives rise to a resonance in the local density of states in the vicinity of the defect.[33,47,50] This state represents the quantization of a semiclassical trajectory in which the electron spirals inwards towards the defect before spiraling back outwards and coupling to a hole that propagates away from the defect.

While it has proven extremely difficult to observe collapse states in relativistic systems, the realization of these states is much easier in graphene. As the Fermi velocity of graphene is much smaller than the speed of light, the critical defect charge for graphene on a hexagonal boron nitride substrate is on the order of unity (compared to $Z_c \sim 170$ for relativistic particles in vacuum). Indeed, collapse states were first observed by Wang et al.[47] who created defects with a controllable charge by assembling Ca atom dimers on graphene. For defects containing three or more dimers, they observed the emergence of a peak in the scanning tunneling spectrum near the defect which was interpreted as the signature of a collapse state. Later, a collapse state was also observed near charged vacancies by Mao and coworkers[37]. Wang et al.[33] were able to induce a so-called frustrated supercritical collapse state in graphene by assembling a chain of molecules with subcritical charges.

    In this paper, we study the behaviour of charged defects in twisted bilayer graphene near the magic angle. As the Fermi velocity approaches zero near the magic angle, one might naively expect that the critical defect charge $Z_c = \epsilon v_F/2$ also becomes very small, suggesting that even a defect with a small charge could induce a collapse state in this system. To address this question, we carry out atomistic tight-binding calculations of tBLG with a charged defect near the magic angle. The defect is modelled as a point charge which induces a Coulomb potential that acts on the electrons in tBLG. We observe that the charged defect gives rise to significant changes in the flat bands that depend sensitively on the position of the defect in the moiré unit cell. However, we do not observe signatures of a collapse state in the flat band manifold. To understand this finding, we analyze the evolution of the collapse state of a graphene monolayer as the hopping to the second (twisted) graphene layer is turned on. We find that the collapse states on both monolayers hybridize as a result of interlayer hopping.

Surprisingly, the hybridized states remain highly localized in tBLG (which is why they do not affect the flat band manifold), but their splitting destroys the associated peak in the local density of states. These predictions can be tested in scanning tunnelling spectroscopy experiments.

## 2. Methods

We studied graphene supercells and commensurate tBLG moiré unit cells containing a single charged defect modelled as a point charge. Because of periodic boundary conditions, this means that a periodic superlattice of charged defects is formed. The moiré unit cells were constructed by rotating the top graphene sheet of an AA-stacked bilayer anticlockwise around an axis perpendicular to the graphene sheets that intersects a carbon atom in each layer. The moiré lattice vectors $(\boldsymbol{t_1}, \boldsymbol{t_2})$ of tBLG are given by $\boldsymbol{t_1} = n\boldsymbol{a_1} + m\boldsymbol{a_2}$; $\boldsymbol{t_2} = -m\boldsymbol{a_1} + (n+m)\boldsymbol{a_2}$, where $\boldsymbol{a_1} = a/2(\sqrt{3}, -1)$ and $\boldsymbol{a_2} = a/2(\sqrt{3}, 1)$ are the graphene lattice vectors (with $a$ = 2.46 Å denoting the graphene lattice constant) and $n$ and $m$ are integers. To describe atomic relaxations in tBLG,[51] we employ the approximation from Refs. [30,31,52,53] where the interlayer distance $d$ is expressed as

$$d(\boldsymbol{\delta}) = d_0 + 2d_1 \sum_{i=1}^{3} \cos{(\boldsymbol{a_i} \cdot \boldsymbol{\delta})}, \qquad (1)$$

where $\boldsymbol{a_3} = -\boldsymbol{a_1} - \boldsymbol{a_2}$ and $\boldsymbol{\delta}$ denotes in-plane atomic position relative to the center of the AA region. Moreover, $d_0 = 1/3\,(d_{AA} + 2d_{AB})$ and $d_1 = 1/9\,(d_{AA} - 2d_{AB})$ with $d_{AB}$ = 3.35 Å being the interlayer distance in the AB-stacked regions and $d_{AA}$ = 3.60 Å being the maximum interlayer spacing in AA regions.

To calculate the electronic structure, we employed an atomistic tight-binding approach.[54] The tight-binding Hamiltonian is given by

$$\hat{H} = \sum_i \varepsilon_i \hat{c}_i^\dagger \hat{c}_i + \sum_{ij} t(\boldsymbol{r}_i - \boldsymbol{r}_j) \hat{c}_i^\dagger \hat{c}_j + h.c., \qquad (2)$$

where $\varepsilon_i$ denotes the on-site energy of the p$_z$-orbital on atom $i$ and $\hat{c}_i^\dagger (\hat{c}_i)$ creates (annihilates) an electron in this orbital. Here, the spin label is left implicit. The hopping parameter between atoms $i$ and $j$ is denoted by $t(\boldsymbol{r}_i - \boldsymbol{r}_j)$ and is given by the Slater-Koster approximation[55]

$$t(\boldsymbol{r}) = V_{pp\sigma}(\boldsymbol{r}) \left( \frac{\boldsymbol{r} \cdot \boldsymbol{e_z}}{|\boldsymbol{r}|} \right)^2 + V_{pp\pi}(\boldsymbol{r}) \left( 1 - \left[ \frac{\boldsymbol{r} \cdot \boldsymbol{e_z}}{|\boldsymbol{r}|} \right]^2 \right), \qquad (3)$$

where $V_{pp\sigma}(\boldsymbol{r}) = V_{pp\sigma}^0 \exp\{q_\sigma(1 - \frac{|\boldsymbol{r}|}{d_{AB}})\} \Theta(R_c - |\boldsymbol{r}|)$ and $V_{pp\pi}(\boldsymbol{r}) = V_{pp\pi}^0 \exp\{q_\pi(1 - \frac{|\boldsymbol{r}|}{a})\} \Theta(R_c - |\boldsymbol{r}|)$. The parameters we used are as follows: $V_{pp\sigma}^0$ = 0.48 eV, V$_{pp\pi}$ = −2.7 eV, $a$ = 1.42 Å (carbon-carbon bond length), $q_\sigma$ = 7.43 and $q_\pi$ = 3.14.[30,31] Hopping parameters between carbon atoms whose distance is larger than the cutoff $R_c$ = 10 Å were neglected.[56]

The onsite energy of an electron in a p$_z$-orbital in the presence of a charged defect with charge $q$ is given by

$$\varepsilon_i = -\frac{e\tilde{q}}{4\pi\epsilon_0 |\boldsymbol{r}_i - \boldsymbol{r}_0|}, \qquad (4)$$

where $\boldsymbol{r}_i$ is the position of atom $i$, $\boldsymbol{r}_0$ is the location of the charged defect, $\tilde{q} = q/\epsilon$ is the effective (screened) charge of the defect and $e$ is the proton charge. Here, we assume that the potential created by the defect is Coulomb-like, with all screening processes (both external and internal) being captured through an effective dielectric constant $\epsilon$. The typical value of $\epsilon$ for tBLG encapsulated in h-BN sheets is ~10.[57,58] A more rigorous approach would be to calculate the screened defect potential using the random phase approximation (RPA)[58] or to use a screened interaction that models the presence of metallic gates.[58,59]

In principle, charged defects can reside at a large variety of inequivalent positions in the moiré unit cell of tBLG. They can intercalate between the layers[41] or adsorb on the surface. Here, we focus on the intercalated charged defects and study the electronic structure for different high-symmetry positions in the moiré unit cell (we have found that the electronic structure of adsorbed and intercalated charged defects is very similar, see Supporting

Information **Figure S1**). In particular, we consider the defect to be located at the center of the AA, AB and bridge regions (br), shown schematically in **Figure 1**. We focus on twist angles that are sufficiently small such that the defect potential decays significantly within a single moiré unit cell.

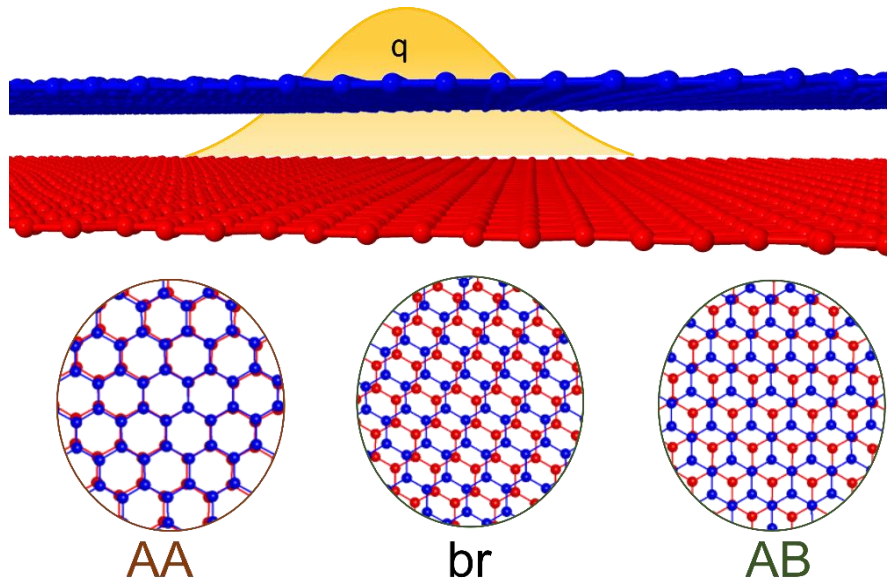

**Figure 1** Top: Schematic depiction of tBLG with the potential due to a charged defect (with charge q). Bottom: different high-symmetry stacking regions (AA, br, AB) that exist in the moiré unit cell of tBLG.

## 3. RESULTS and DISCUSSION

### Effect of charged defects on the flat bands

We first discuss the dependence of the flat band manifold on the position of the charged defect in the moiré unit cell. If one places the charged intercalant at the center of the AA region, no symmetries are broken. Therefore, the band structure still exhibits a Dirac cone at the K and K' points, as shown in **Figure 2** (a) – (c) for a twist angle of 1.54° for a negatively charged defect. As the magnitude of the effective defect charge increases, the Dirac cone at the K-point shifts up in energy relative to the states at $\Gamma$. Conversely, for a positively charged defect, the states at the K-point shift downwards relative to the states at the $\Gamma$-point (see

Supporting Information **Figure S2**). For $\tilde{q} = -0.1$, the low-energy conduction bands are extremely flat. For $\tilde{q} = -0.15$, the dispersion of the lowest conduction band is qualitatively different and exhibits a negative curvature near $\Gamma$, similar to the highest energy valence bands near $\Gamma$. Similar observations were reported in Ref. [41] for a Li atom intercalated in tBLG for twist angles of 7.34° and 2.45°.

Interestingly, these band distortions are remarkably similar to those obtained from Hartree theory calculations of doped tBLG without charged defects. [8,10,60,61,19,62,32] The flat band states near K and K' are localized in the AA regions of the moiré unit cell.[63] Therefore, when electrons or holes are added into the lowest-energy flat band states, the charge density in the AA regions changes, creating a localized charge density similar to that of a charged defect. This causes a large change in the Hartree potential. When electrons are added (removed), the Hartree potential is positive (negative) in the AA regions. As the states at the edge of the Brillouin zone (e.g., at K/K' and M) are localized on the AA regions [see Supporting Information **Figure S3** (b) and (d)], these states are pushed to higher (lower) energies. In contrast, states near the center of the Brillouin zone are localized on the AB/BA regions, and their energies do not change significantly upon doping. This difference in the response of the states at the center and the corners of the Brillouin zone explains the strong band deformations that are observed upon doping or adding a charged defect in the AA regions.[19,60,61]

In contrast, when the defect is placed at the center of the AB regions of the moiré unit cell, a gap opens at the K and K' points, and the Dirac cone is destroyed (see **Figure 2** (d) – (f) at a twist angle of 1.54°). With increasing defect charge, this gap widens. We can understand this gap opening from an analysis of the flat-band Wannier functions.[12,53,64] These Wannier functions are centered on the AB and BA regions of the moiré unit cell, which can be thought of as the two sublattices of the moiré flat bands, similar to the *A* and *B* sublattices of graphene. Therefore, upon placing a defect on one of the sublattices, the sublattice symmetry is broken which protects the Dirac cone at K/K' and a gap is opened.

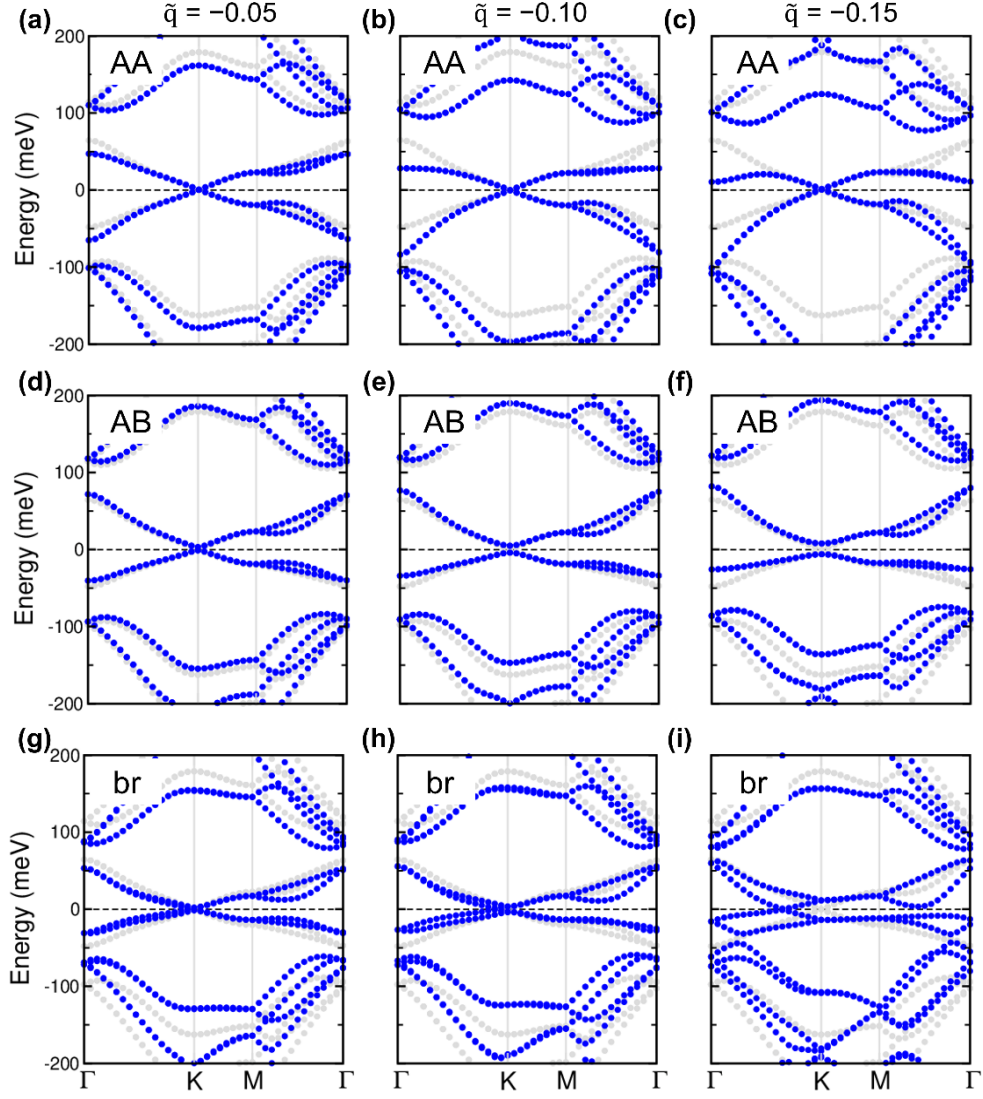

**Figure 2** Electronic band structure of twisted bilayer graphene (with a twist angle of 1.54°) with an intercalated charged defect at the center of the (a)–(c) AA, (d)-(f) AB, and (f)-(h) bridge (br) regions. For each region three different $\tilde{q}$ values (-0.05, -0.10, and -0.15) are considered and, for reference, the band structure for $\tilde{q} = 0$ is also included in light grey color. The zero of energy (indicated by the horizontal black dashed lines) is set to the energy of the Dirac point or the center of the band gap in each case.

In addition to the gap opening, we also observe some distortions of the flat bands. As $\tilde{q}$ increases, the states near Γ increase in energy relative to the states at K/K'. As the states near Γ are localized in the AB and BA regions of the moiré unit cell, the potential of the charged defect couples more strongly to these states than to the states at K and K'. It is interesting to note that the band deformations are less pronounced compared to the case when the defect is placed in the AA regions. This is a consequence of the lower degree of localization of the states near Γ compared to those at K/K'.

Finally, tBLG with a defect in the center of the bridge regions does not exhibit Dirac points at K and K', see Figure 2 (g)-(h). However, the system is metallic as a band from the valence manifold of tBLG crosses a band from the conduction manifold along the Γ to K line, see Figure 2 (h).

Another variable which can be tuned in tBLG is the twist angle. In **Figure 3**, we show the evolution of the band structure of undoped tBLG with a charged defect with $\tilde{q} = -0.1$ in the AA region as the twist angle is reduced (see Supporting Information **Figure S4** for AB and br regions). As discussed above, this system exhibits extremely flat conduction bands at low energies for a twist angle of 1.54°, see **Figure 3** (a). Upon decreasing the twist angle towards the magic angle, the effect of the defect becomes more pronounced and the states at K are pushed up so much that the sign of the curvature of the lowest conduction band near Γ changes, see **Figure 3** (b). Very close to the magic angle, at a twist angle of 1.05°, the flat bands are again strongly distorted by the defect potential, and the gap between the flat bands and the remote bands becomes very small, see **Figure 3** (c). Again, these findings are reminiscent of the band structure of doped tBLG as the twist angle approaches the magic angle.[5,65]

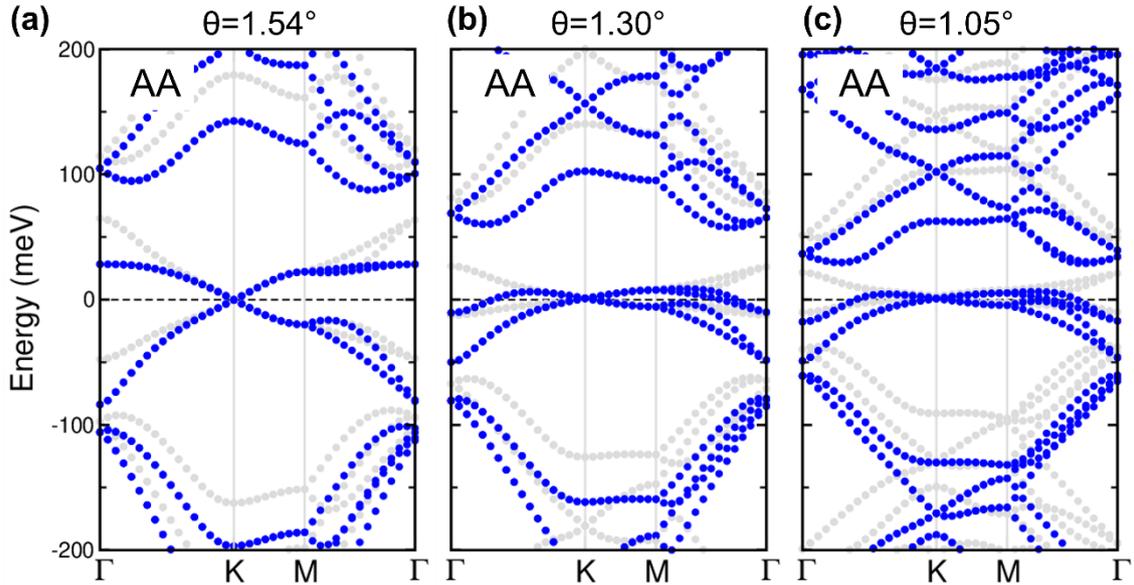

**Figure 3** Electronic band structures of twisted bilayer graphene with an intercalated charged defect ($\tilde{q} = -0.1$) at the AA region for three twist angles: (a) 1.54°, (b) 1.30°, and (c) 1.05°. The band structure without the defect is shown in light grey for comparison. The zero of energy (indicated by the horizontal black dashed lines) is defined as the Dirac point energy.

**Collapse states in graphene and their fate in tBLG**

Before discussing the existence of a collapse state in tBLG, we first revisit the collapse states in monolayer graphene from a band structure perspective.

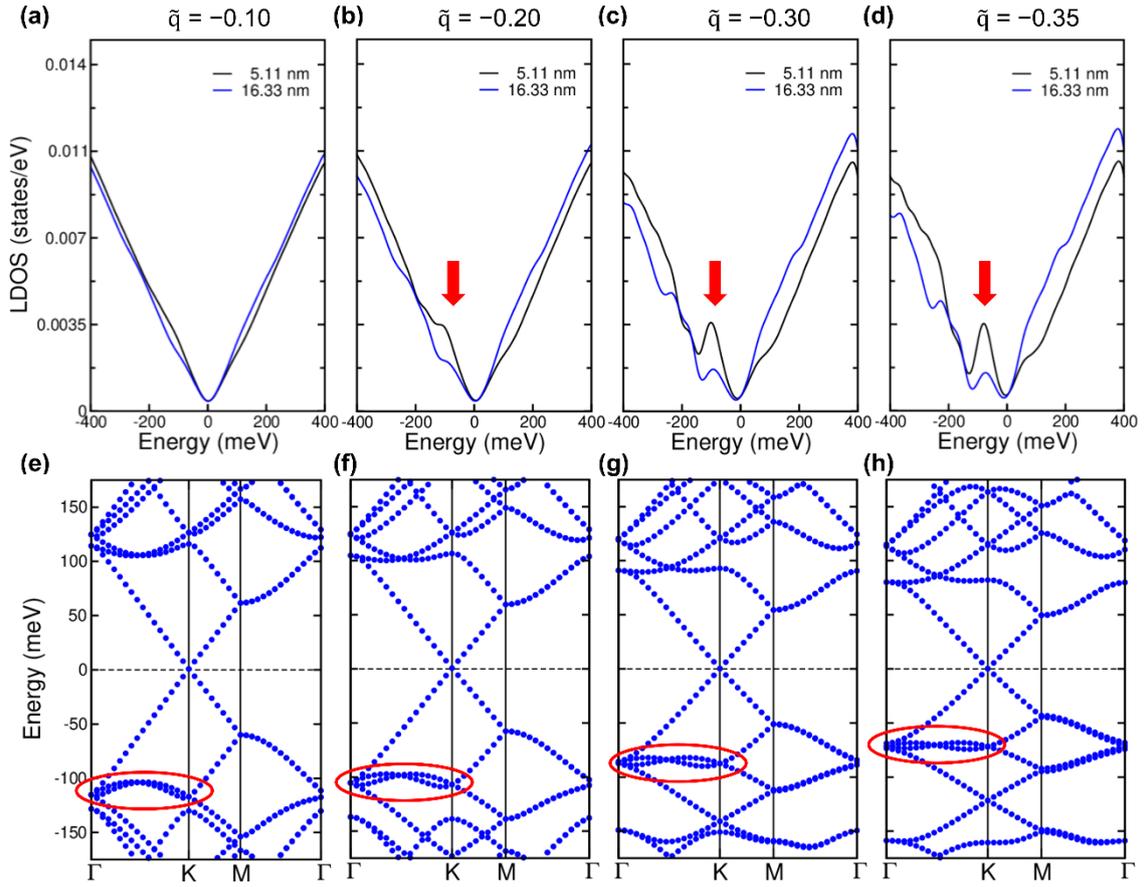

**Figure 4** (a)–(d) Local density of states of an 80 × 80 monolayer graphene supercell with a charged adsorbate for two distances from the defect and for four different values of the effective defect charge. Red arrows indicate the peak due to the collapse state. (e)–(h) Corresponding electronic band structures with the bands that give rise to the collapse state circled in red. The zero of energy is referenced to the energy of the Dirac point.

In **Figure 4** (a)–(d), we show the LDOS of an 80 × 80 graphene supercell with a charged adsorbate as function of energy for two distances (5.11 nm and 16.33 nm) from the defect for different effective defect charges. The charged defect is placed directly above a carbon atom in the center of supercell at a distance of 0.5Å above the graphene plane. The LDOS of graphene has a characteristic V-shape at low energies.[66] For small effective defect charges (see **Figure 4** (a) for an effective charge of $\tilde{q} = -0.1$), the LDOS is not significantly altered by the presence of the defect. As the effective defect charge is increased, an additional peak corresponding to a collapse state emerges in the LDOS, as highlighted by the red arrows. For a negatively charged defect, the peak appears on the valence side at an energy of approximately 100 meV below the Dirac point. Moreover, at slightly more negative energies than the peak, a pronounced dip in the LDOS can be observed, see **Figure 4** (c). At energies above the Dirac point, the LDOS near the defect is reduced compared to its value far away from the defect and also exhibits a clear electron-hole asymmetry.[67] For a positively charged defect, we find a similar behaviour but with the important difference that the collapse state peak is located at energies above the Dirac point (see Supporting Information **Figure S5**). These observations are well known from previous studies of charged defects in monolayer graphene.[34,47]

Next, we analyze the origin of the LDOS peak in terms of the band structure of the 80 × 80 supercell with a charged defect. For a periodic superlattice of charged defects on graphene, the band structure is an experimental observable and can be accessed, for example, through angle-resolved photoemission spectroscopy.[68] In **Figure 4** (e)−(h), we show the band structures corresponding to the LDOS plots in **Figure 4** (a)−(d). We note that the details of the band structure depend on the choice of supercell. For the 80 × 80 supercell that we studied, the Dirac cone is mapped onto the K-point of the mini Brillouin zone. For small effective defect charges, such as $\tilde{q} = -0.1$, we do not observe any significant changes in the band structure. As the magnitude of the effective defect charge is increased to $\tilde{q} = -0.2$ or -0.3, the bands with energies of approximately −100 meV flatten between K and Γ (indicated by the red circles), giving rise to the peak in the LDOS. This critical value of the defect charge is consistent with the analytical result $Z_c = \frac{v_F}{2} \approx 0.25$ for a graphene sheet in vacuum. As the defect charge increases even more, these bands become even flatter and move to higher energies. Interestingly, a second Dirac cone emerges below the flat bands which gives rise to the dip in the LDOS beyond the collapse state peak. For a positively charged defect, this secondary Dirac cone emerges above the flat collapse-state bands (see Supporting Information **Figure S5**).

Next, we study the fate of the collapse state in tBLG. **Figure 5** shows the evolution of the band structure of 1.54° tBLG with a charged intercalant with $\tilde{q} = -0.30$ in the AA region (results for charged defects in the AB and br regions are shown in the Supporting Information, see **Figure S6**) as the hopping between the two graphene layers is gradually "turned on" (see Supporting Information **Figure S7** for the same analysis for tBLG without a charged defect). This is achieved by scaling the Slater-Koster parameter $V_{pp\sigma}^0$, which controls the strength of interlayer hopping, by a parameter $\alpha$ that ranges from 0 to 1. For $\alpha$ = 0, the band structure is very similar to that of the 80 × 80 graphene supercell, see **Figure 5**(a), with only small differences arising from the (non-zero, but small) $V_{pp\pi}^0$ hoppings between the layers. As $\alpha$ increases, the band structure undergoes significant changes. In particular, the Fermi velocity decreases and the flat bands corresponding to the collapse state (at approximately -190 meV in **Figure 5**(a) and circled in green) split and become dispersive in the regions between K and $\Gamma$. As the inter-layer hopping is further increased, the splitting of the collapse state bands increases and the lower pair crosses a deeper-lying band (circled in orange in **Figure 5**(a)-(d)).

To further understand the interplay between collapse states and the moiré potential, we analyze the wavefunctions of the relevant states at $\Gamma$ for several values of $\alpha$ (see **Figure 5** bottom panel). Inspection of the wavefunctions for $\alpha = 0$ reveals that the collapse states are strongly localized in the vicinity of the charged defect, while the states below the collapse states are delocalized over the AB and br regions. For $\alpha = 0.25$, the collapse state bands split, but the in-plane localization of the corresponding wavefunctions does not change significantly compared to $\alpha = 0$. However, interlayer hopping results in a coupling of the monolayer collapse states which gives rise to the observed energy splitting. As $\alpha$ is further increased, the splitting of the hybridized collapse states further increases and the lower-lying collapse-state bands eventually cross the deeper-lying bands (see **Figure 5**(c) and (d)). While the formation of hybridized collapse states gives rise to a significant energy splitting at $\Gamma$, symmetry requires that the collapse state bands remain degenerate at K and K'. As a consequence, the localized collapse state bands acquire a significant dispersion along the $\Gamma$ to K/K' direction.

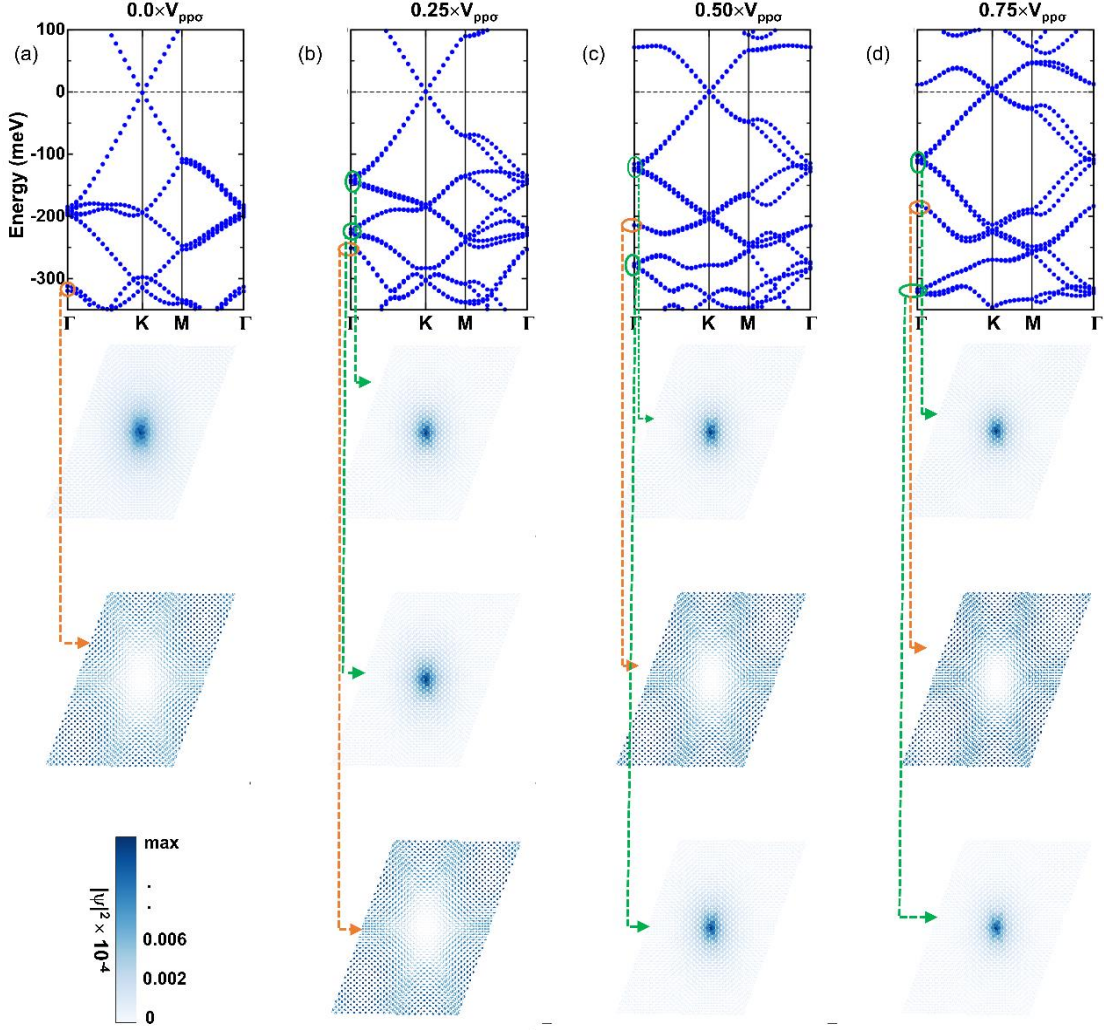

**Figure 5** (Top panel) Evolution of the electronic band structure of twisted bilayer graphene (twist angle of 1.54°) with an intercalated charged defect ($\tilde{q}$ = -0.30) at the AA region as the interlayer hopping (described by $V_{pp\sigma}$ which is multiplied by a scaling factor $\alpha$ ranging from 0 to 1) is turned on. (Bottom panels) The square modulus of the wavefunctions enclosed in the circle indicated in the top panel. Note that the square modulus of all the degenerate states in each circle are summed up. Green circles denote localized collapse states and orange circles denote delocalized states of remote bands.

Finally, **Figure 6**(a) shows the local density of states of twisted bilayer graphene with and without a charged defect at a distance of 2.93 nm, along the diagonal of the cell, away from the defect site. Even without a charged defect ($\tilde{q} = 0$), the LDOS exhibit a series of peaks arising from van Hove singularities of the different minibands. When the defect charge is increased to $\tilde{q} = -0.1$, we observe that features in the LDOS are shifted and peak intensities are changed. For example, the peak near +25 meV significantly increases because of band flattening induced by the defect. For a defect charge which is above the critical value for monolayer graphene ($\tilde{q} = -0.3$), more dramatic changes in the LDOS can be observed. In particular, the peaks at positive energies become more prominent and shift to lower energies. Importantly, we do not observe a signature of collapse states in the LDOS. Such signatures are expected to occur at negative energies: as shown in Figure 5 (d), the higher-lying set of bands with collapse state character have an energy of -100 meV at Γ, but there is no corresponding peak in the LDOS. As discussed above, this is a consequence of the dispersion acquired by the collapse state bands arising from interlayer hybridization.

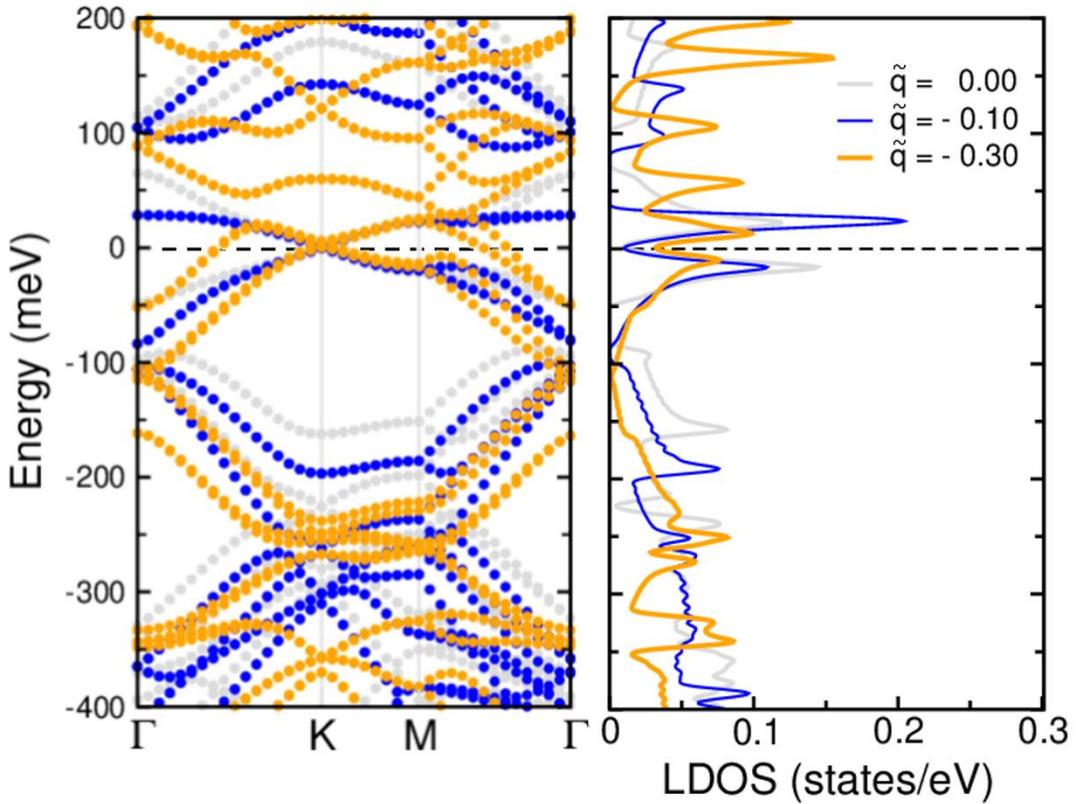

**Figure 6** The local densities of states (LDOS) for tBLG with 1.54° twist angle for effective charges of 0.0, -0.10 and -0.30, intercalated at the AA region. The LDOS is calculated at a distance 2.93 nm away from the defect using 31 × 31 × 1 regular k-grid. The zero of energy is referenced to the energy of the Dirac point.

## 4. Conclusion

We have studied the electronic structure of twisted bilayer graphene with a charged defect near the magic angle. In contrast to the naïve expectation that the strongly reduced

Fermi velocity in tBLG results in a very small value of the critical defect charge for inducing collapse states in the flat band manifold, we find that the charged defect only induces deformations of the flat bands which depend sensitively on the position of the defect in the moiré unit cell. For example, we observe an additional flattening of the flat bands that is reminiscent of the band structure of the doped system without defects for charged defects at the center of AA regions. This finding suggests that it is possible to control electronic phases of twisted bilayer graphene through defect engineering. We also analyze the fate of the monolayer collapse state in twisted bilayer graphene and find that the monolayer collapse states hybridize with an associated energy splitting at the $\Gamma$−point of the moiré Brillouin zone, but remain highly localized in the in-plane directions. Despite their localized character, the collapse state does not induce a peak in the local density of states as a result of the significant dispersion of this state throughout the Brillouin zone. These predictions can be tested in scanning tunneling microscopy and spectroscopy experiments.

## 5. DATA AVAILABILITY

The data to reproduce the plots and findings within this paper are available from the corresponding author upon reasonable request.

## 6. Acknowledgments

M.S.R and A.K acknowledge financial support by Deutsche Forschungsgemeinschaft (DFG, German Research Foundation) within SFB1415 project number 417590517, RTG2247 (QM3), and the association with the SPP2244 (2DMP), and the high-performance computing center of ZIH Dresden for computational resources. M.S.R acknowledges financial support by project SMART, financed by the Volkswagen Foundation as part of the program "Niedersächsisches Vorab - Digitalisierung in den Naturwissenschaften. ZG was supported through a studentship in the Centre for Doctoral Training on Theory and Simulation of Materials at Imperial College London funded by the EPSRC (EP/L015579/1). Z.G. acknowledges support by EC-FET European Graphene Flagship Core3 Project, EPSRC grants EP/S030719/1 and EP/V007033/1, and the Lloyd Register Foundation Nanotechnology Grant. J.L and A. A.

M. acknowledge funding from EPSRC grant EP/S025324/1 and the Thomas Young Centre under grant number TYC-101. We thank V. Falko and V. Bacic for stimulating discussions.

## 7. Author contributions

J.L conceived the idea; M.S.R performed most of the simulations under the guidance of Z.G. All authors contributed to the discussion and analyzed the results and wrote this manuscript together.

## 8. COMPETING INTERESTS

The authors declare no competing financial or non-financial interests.